\documentclass[twocolumn]{aastex62}

\graphicspath{{./}{figures/}}

\shorttitle{Double white dwarfs in the Local Group with LISA}
\shortauthors{Korol et al.}

\begin{document}

\title{Detectability of double white dwarfs in the Local Group with LISA}

\correspondingauthor{Valeriya Korol}
\email{korol@strw.leidenuniv.nl}

\author[0000-0002-0786-7307]{Valeria Korol}
\author{Orlin Koop}
\author{Elena M. Rossi}
\affil{Leiden Observatory \\
Neils Bohrweg 2, 2333 CA, Leiden, the Netherlands}

\begin{abstract}

Detached double white dwarf (DWD) binaries are one of the main science case for the Laser Interferometer Space Antenna (LISA). As the most numerous LISA sources, they will provide important contributions towards understanding binary evolution, Supernovae Type Ia (SNIa) formation channels and the structure of the Milky Way. So far only detection prospects for the Milky Way have been computed. In this letter we show that LISA has the potential to detect DWDs in neighboring galaxies up to the border of the Local Group. In particular, we compute quantitative estimates for the number of detections in M31. We expect between a dozen to several tens of DWDs above the nominal detection threshold, for a mission duration between 4 and 10$\,$yr. We show that extra-galactic DWDs detectable by LISA are those with the shortest orbital periods and with the highest chirp masses, that are candidates SNIa progenitor, virtually undetectable at those distances in optical. This implies that LISA will be the only instrument able to provide SNIa merger rates across the Local Group.

\end{abstract}


\section{Introduction} \label{sec:intro}

Detached DWD binaries with orbital periods $< 1\,$h will be important GW sources for the LISA mission in many ways \citep{ama17}. Firstly, DWDs are guaranteed LISA sources. A number of short period DWDs have already been identified at optical wavelengths \citep[e.g.,][]{kup18}. Those with strongest signals can be used as calibration sources as they will be detectable already after one week of observations; over time their signal will increase improving the accuracy with which these sources can be used to monitor data quality as new data are acquired \citep[][]{lit18}. Secondly, DWDs will be the most numerous LISA sources. The total number of expected detections exceeds $10^5$ \citep[e.g.,][]{nel01,rui10,mar11,kor17}. Thus, for the first time LISA will provide a sizeable sample of short period DWD binaries to test binary formation theories and validate SNIa formation channels \citep[e.g.,][]{nel01,nel04,reb18}. Moreover, such a large number of individually resolved sources spread all over the Galaxy will allow us to map the Milky Way in GWs and precisely measure its structural parameters like scale radii of the  bulge and the disc \citep{ada12,kor18}. When combining GW and optical measurements for DWDs with optical counterparts we will also be able to derive the mass of the bulge and the disc component of the Galaxy \citep{kor18}. Finally, these binaries are so common in the Milky Way that their unresolved signals will form a background for the LISA mission \citep[e.g.,][]{rob17a}. This background contains information on the overall population of DWDs in the Milky Way and can be also used to derive the Milky Way's parameters, like the disc scale height \citep{ben06}. 

Previously the detectability of DWDs has been exclusively assessed in the Milky Way, while extra-galactic DWDs were only considered as contribution to the background noise \citep[e.g.,][]{kos98,far03}. In this letter we focus for the first time on the properties of extra-galactic DWDs that can be resolved by LISA in the Local Group, and especially in the Large and Small Magellanic clouds (LMC and SMC), and M31 (the Andromeda galaxy). We show that LISA will detect binaries with the shortest periods and highest total masses, and therefore double degenerate SNIa progenitors. As discussed in \citet{reb18}, these are difficult to find with optical telescopes in the Milky Way. Essentially, they are too faint to be identified from H$\alpha$ double-lined profiles in spectra and their eclipses are too short when considering the typical cadence of observations of optical sky surveys, like {\sl Gaia}. Therefore LISA might be the best tool to allow statistical studies of these systems.

In this letter, we forecast the parameter space of DWDs accessible through GW observation located at the distance of SMC, LMC and M31 (Section \ref{sec:2}).  We use a synthetic population to quantify the number of detection for M31 (Section \ref{sec:3}). In Section \ref{sec:4} we present our conclusions.


\section{Maximal distance} \label{sec:2}

In this section we consider an illustrative example of a monochromatic DWD binary, i.e. a binary whose orbital period decay due to GW emission is too small to be measured during the mission lifetime. A monochromatic assumption is justified when interested in the signal-to-noise ratio (SNR) only, as in this letter. However, the measurement of the orbital period decay is essential to recover the binary chirp mass and the distance from GW data. For a monochromatic source SNR can be estimated as \citep[e.g.,][]{mag08}
\begin{equation} \label{eq:snr}
{\rm SNR} = {\cal A} \ F(\iota, \theta, \phi, \psi)\  \sqrt{\frac{T_{\rm obs}}{S_{\rm n}(f)}},
\end{equation}
where ${\cal A}$ is the amplitude of the GW signal, $F(\iota, \theta, \phi, \psi)$ is a function that accounts for the instrument response to the binary inclination $\iota$, sky position $(\theta, \phi)$ with respect to the detector and polarization angle $\psi$ averaged over one LISA orbit \citep[][equation~(42)]{cor03}, $T_{\rm obs}$ is the observation time and $S_{\rm n}(f)$ is the noise spectral density at the binary frequency $f=2/P$, with $P$ being the binary orbital period. The amplitude can be computed using the quadrupole approximation as
\begin{equation}\label{eq:amp}
{\cal A} = \frac{4(G{\cal M})^{5/3}(\pi f)^{2/3}}{c^4d},
\end{equation}
where ${\cal M}=(m_1 m_2)^{3/5}/(m_1 + m_2)^{1/5}$ is the chirp mass and $d$ is the distance \citep[e.g.,][]{mag08}. We draw $\psi$ randomly from a flat distribution between $[0,\pi]$. We adopt the sky- and inclination-averaged noise curve, $S_{\rm n}(f)$, corresponding to the LISA mission design accepted by ESA with nominal and extended mission duration of $4\,$yr and $10\,$yr respectively \citep{ama17}. Finally, we average the SNR over the inclination angle $\iota$ and the position in the sky $(\theta, \phi)$. Equation~(\ref{eq:amp}) shows that the strength of the signal mainly depends on three intrinsic binary parameters: $f ($or $P), {\cal M}$ and $d$. Thus, we study binary detectability of DWDs with LISA as a function of these parameters.
\begin{figure}
	\includegraphics[width=0.5\textwidth]{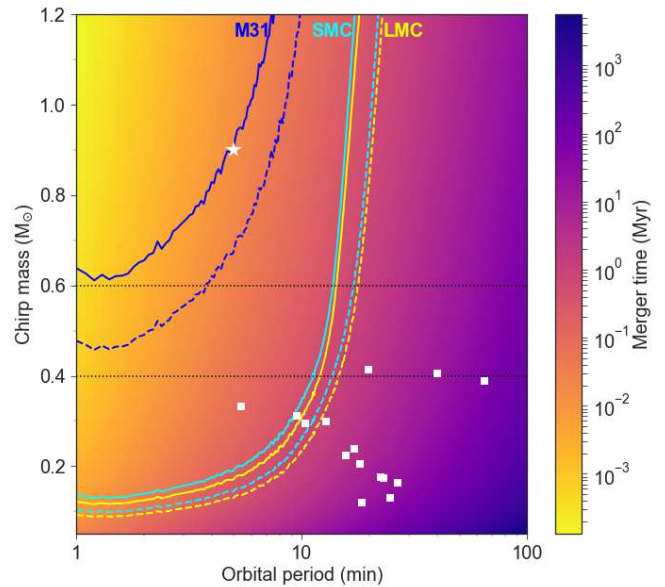}
    \caption{Curves of sky- and inclination-averaged SNR=7 in orbital period-chirp mass space evaluated at the distances of the LMC, SMC and M31 for $4\,$yr (solid) and $10\,$yr (dashed) mission lifetime. The color represents the merger time. White squares are known Galactic GW sources (DWDs and AM CVns) and the white star is our test binary.}
    \label{fig:1}
\end{figure}
\begin{figure*}
	\centering
    \includegraphics[width=0.86\textwidth]{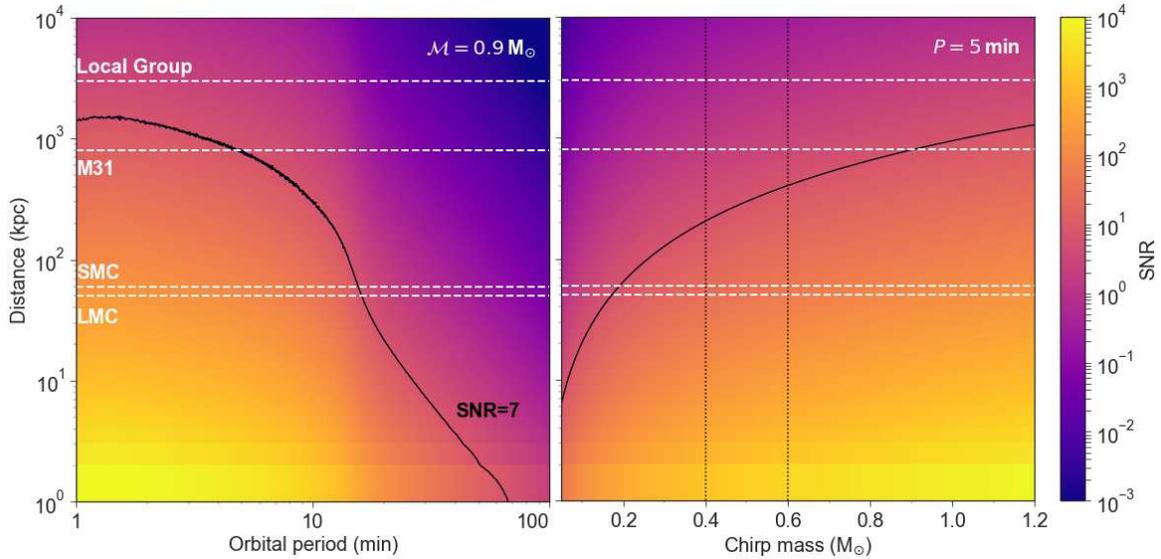}
    \caption{SNR for the test DWD in the distance-orbital period for ${\cal M}=0.9\,$M$_{\odot}$ (left) and distance-chirp mass for $P=5\,$min (right) space after nominal $4\,$yr of observations. The black solid line represents the iso-SNR contour of 7. Dashed horizontal lines mark the distance of the LMC, SMC, M31 and the border of the Local Group.}
    \label{fig:2}
\end{figure*}

In Figure~\ref{fig:1} we plot in $P-{\cal M}$ space the sky- and inclination-averaged curves of SNR=7 at the distance of the LMC (yellow), SMC (cyan) and M31 (blue) for 4 yr (solid) and 10 yr (dashed) mission lifetimes. Thus, the areas above the curves delimit the parameter space detectable by LISA in these galaxies. At the distance of the LMC and SMC, LISA will be sensitive to DWDs with any chirp mass when the orbital period $<20\,$min. Whereas at the distance of M31 LISA will be sensitive only to binaries with ${\cal M}>0.5\,$M$_{\odot}$ and $P<10\,$min. The color contour shows the binary merger time:
\begin{equation} \label{eq:tau}
\tau \simeq 1 \,{\rm Myr} \  \left( \frac{P}{12\,{\rm min}} \right)^{8/3} \left( \frac{{\cal M}}{0.3\,{\rm M}_{\odot}} \right)^{-5/3}. 
\end{equation}
Thus, Figure~\ref{fig:1} revels that DWDs accessible to LISA in the three considered galaxies will merge in $<1\,$Myr. The two horizontal lines at ${\cal M} = 0.6, 0.4\,$M$_{\odot}$ correspond to binaries with a total mass equal to the Chandrasekhar mass when adopting equal mass and unequal mass ($m_2/m_1 = 1.8$) binary components respectively. They are lower bound of the parameter space corresponding to double degenerate SNIa progenitors. Finally, the white squares are known Galactic GW sources (DWDs and AM CVns) from \citet[][]{kup18} showing the part of parameter space explored so far at optical wavelengths. In particular, those with the shortest periods (HM Cnc, V407 Vul, ES Cet and SDSS J0651) could be detected by LISA if placed at the distance of LMC and SMC. On the other hand, the parameter space accessible in M31 is currently unprobed.

Next, we consider a test binary with ${\cal M}=0.9\,$M$_{\odot}$ and $P=5\,$min (white star in Figure~\ref{fig:1}), such that it could be detected in M31. LISA will detected a handful of binaries with the same parameters in the Milky Way \citep{reb18}. Figure~\ref{fig:2} represents the sky- and inclination-averaged SNR in the distance-orbital period parameter space for fixed  ${\cal M}=0.9\,$M$_{\odot}$ (left panel) and in distance-chirp mass parameter space for fixed $P=5\,$min (right panel). The black solid line shows the LISA detection threshold of 7. The area below the curve represents the parameter space detectable by LISA and shows that LISA has the potential to detect our test source almost up to the border of the Local Group.

\section{DWD detections in Andromeda} \label{sec:3}
\begin{figure*}
	\centering
    \includegraphics[width=0.85\textwidth]{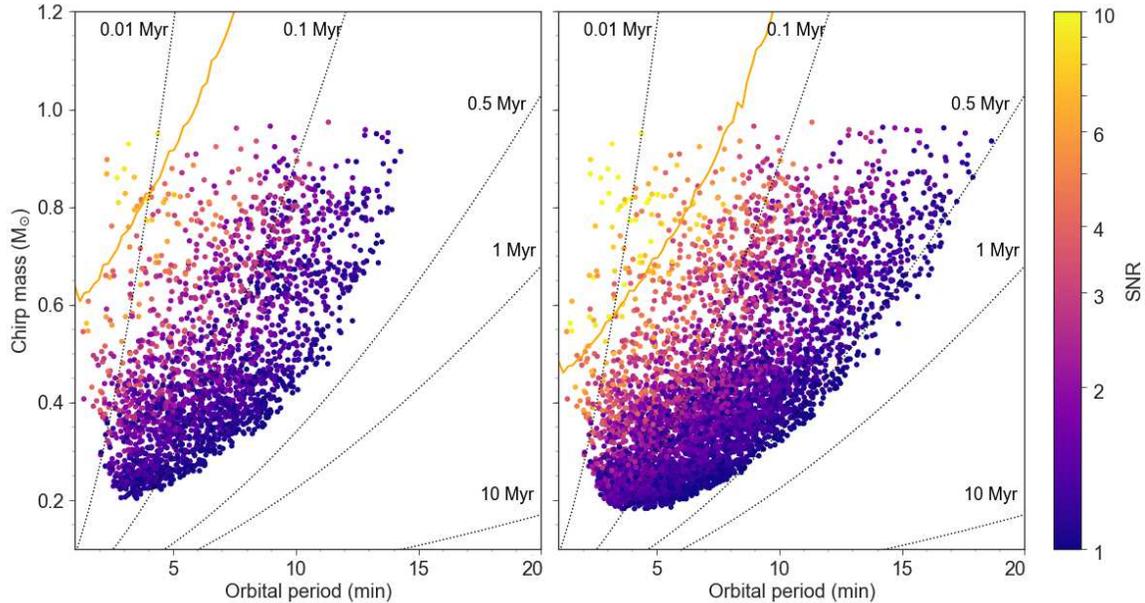}
    \caption{DWDs in M31: after $4\,$yr (left panel) and after $10\,$yr (right panel) observation time with LISA. Orange solid line represents sky- and inclination-averaged LISA detection threshold of 7. Black dotted lines are iso-merger contours.}
    \label{fig:3}
\end{figure*}

Here we address quantitative estimates for the number of detections in the Andromeda galaxy. Properties of extra-galactic DWDs are not known, since not a single DWD has been observed outside our Galaxy. However, because M31 is a spiral galaxy similar to the Milky Way, we can extrapolate the properties of the Galactic population of DWDs to that of Andromeda. Practically, we assume the same distributions for binary initial parameters (like the initial mass function, orbital period and eccentricity distributions), binary fraction and metallicity as in the Milky Way \citep[for details see][]{too12,kor17}, and we double the total stellar mass. For all binaries we assign galactic coordinates $(l,b) = (121, -21)$, $d=800\,$kpc and an inclination angle randomly drawn from a uniform distribution in $\cos \iota$. We compute the SNR as described in Section~\ref{sec:2}. We find 12 (60) binaries with SNR $\ge 7$ in 4 (10) yr of the mission (see Figure~\ref{fig:3}). In particular, the majority are CO+CO and the small fraction are CO+ONe DWDs. Our predictions are likely underestimates for a few reasons. First, because the number of detections heavily depends on the binary inclination angle. For example, when assigning a favorable orientation for detection (i.e. $\iota=0$) to all binaries we find 60 (230) detectable sources in 4 (10) yr of observation. Secondly, we suggest that a targeted search for DWDs at the position and distance of M31 might increase the chance of finding them in the LISA data. Figure~\ref{fig:3} shows the properties of the binaries in M31. All the binaries detectable by LISA (SNR$>7$) will merge in less than $0.1\,$Myr. We also find that the LISA sample is complete for $P<3.5\,$min and ${\cal M} > 0.7\,$M$_{\odot}$. The completeness is $\sim 50$\% for DWDs with $P<4.5\,$min and ${\cal M} > 0.6\,$M$_{\odot}$, and drops to $\sim 10$\% for $P<10\,$min and ${\cal M} > 0.6\,$M$_{\odot}$. Given the completeness and the merger time across the sample we will be directly able to estimate the DWD merger rate in M31. 

\section{Conclusions} \label{sec:4}

In this letter we explored the detectability of DWDs outside the Milky Way. We proved that LISA has the potential to detect binaries in neighboring galaxies: LMC, SMC and M31. We find that in the LMC and SMC LISA can detect any binary with $P< 20\,$min, while in M31 LISA will be sensitive to DWDs with $P<10\,$min and ${\cal M} > 0.5\,$M$_{\odot}$. Using an example DWD with $P=5\,$min and ${\cal M}=0.9\,$M$_{\odot}$, we showed that binaries with such characteristics can be detected up to $1\,$Mpc distance, i.e. within the large volume of the Local Group. In the Andromeda galaxy we found a few, to several tens, of DWDs above the LISA detection threshold for 4 and 10$\,$yr mission. This gives an optimistic prospects for detecting other kind of stellar type GW sources like AM CVns and low-mass X-ray binaries, which will likely have an electromagnetic counterpart. A large fraction of extra-galactic DWDs detectable by LISA will have total mass exceeding the Chandrasekhar mass limit and will merge in less than $1\,$Myr, meaning that LISA has the potential to provide SNIa merger rates across the Local Group.

\acknowledgments
We thank S. Blunt, K. Breivik, W. R. Brown, D. Gerosa, K. Kremer, T. Kupfer, A. Lamberts, S. Larson, T. Littenberg and S. Toonen for useful discussions. EMR, VK acknowledge NWO WARP Program, grant NWO 648.003004 APP-GW.

\end{document}